\documentclass[conference]{IEEEtran}
\IEEEoverridecommandlockouts
\usepackage{cite}
\usepackage{amsmath,amssymb,amsfonts}
\usepackage{algorithmic}
\usepackage{graphicx}
\usepackage{textcomp}
\usepackage{xcolor}
\def\BibTeX{{\rm B\kern-.05em{\sc i\kern-.025em b}\kern-.08em
    T\kern-.1667em\lower.7ex\hbox{E}\kern-.125emX}}

\usepackage{hyperref}

\newif\iflong
\longfalse

\usepackage[inline]{enumitem}
\usepackage{ragged2e}

\usepackage[font=small]{caption}
\usepackage{booktabs}
\usepackage{amsmath}
\usepackage{amsthm}
\usepackage{array}
\usepackage{listings}

\usepackage{multicol,multirow}

\usepackage{tikz}
\usetikzlibrary{shapes,arrows,positioning,calc,fit,intersections,through}

\usepackage{xurl}
\usepackage{xspace}

\usepackage[T1]{fontenc}
\usepackage[scaled=0.81]{beramono}

\usepackage{fontawesome}

\newlength{\twemojiDefaultHeight}
\AtBeginDocument{\setlength{\twemojiDefaultHeight}{\fontcharht\font`X}}
\newcommand{\twemoji}[2][height=\twemojiDefaultHeight]{%
  \csname twemoji #2\endcsname{#1}%
}%
\newcommand{\defineTwemoji}[1]{%
\expandafter\newcommand\csname twemoji #1\endcsname[1]{%
  \includegraphics[##1]{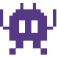}}%
}%
\defineTwemoji{alien monster}

\newcommand{\bip}{{\smaller[0.5]{\textsc{BugsInPy}}}\xspace}

\newcommand{\Py}[1]{\mbox{\lstinline[basicstyle=\ttfamily,language=Python]|#1|}}

\newcommand{\Bash}[1]{\lstinline[basicstyle=\small\ttfamily,language=bash]|#1|}

\newcommand{\fp}{{\smaller \textsc{FauxPy}}\xspace}
\newcommand{\project}[1]{{\smaller[0.5]{\textsf{#1}}}\xspace}
\newcommand{\muse}{{Muse}\xspace}

\usepackage{amsthm}

\definecolor{acol}{RGB}{27,158,119}
\definecolor{bcol}{RGB}{217,95,2}
\definecolor{ccol}{RGB}{117,112,179}
\definecolor{dcol}{RGB}{231,41,138}
\definecolor{ecol}{RGB}{102,166,30}
\definecolor{fcol}{RGB}{230,171,2}
\definecolor{gcol}{RGB}{166,118,29}

\usepackage{pgfkeys,pgfplots,numprint,relsize}
\xspaceaddexceptions{\%}  %
\def\NAN{??}              %
\def\keyfamily{/flpy/}    %

\usepackage{fmtcount}     %

\pgfkeyssetvalue{/flpy/httpie/Tarantula/@X}{1}
\pgfkeyssetvalue{/flpy/httpie/Ochiai/@X}{1}
\pgfkeyssetvalue{/flpy/httpie/DStar/@X}{1}
\pgfkeyssetvalue{/flpy/httpie/Metallaxis/@X}{0}
\pgfkeyssetvalue{/flpy/httpie/Muse/@X}{3}
\pgfkeyssetvalue{/flpy/httpie/PS/@X}{1}
\pgfkeyssetvalue{/flpy/httpie/PS/time}{116.47960531711578}
\pgfkeyssetvalue{/flpy/httpie/ST/@X}{0}
\pgfkeyssetvalue{/flpy/httpie/ST/time}{1.1015335321426392}
\pgfkeyssetvalue{/flpy/httpie/SBFL/time}{9.142874121665955}
\pgfkeyssetvalue{/flpy/httpie/MBFL/time}{645.740253329277}
\pgfkeyssetvalue{/flpy/pandas/Tarantula/@X}{3}
\pgfkeyssetvalue{/flpy/pandas/Ochiai/@X}{3}
\pgfkeyssetvalue{/flpy/pandas/DStar/@X}{3}
\pgfkeyssetvalue{/flpy/pandas/Metallaxis/@X}{2}
\pgfkeyssetvalue{/flpy/pandas/Muse/@X}{1}
\pgfkeyssetvalue{/flpy/pandas/PS/@X}{2}
\pgfkeyssetvalue{/flpy/pandas/PS/time}{29652.790944417316}
\pgfkeyssetvalue{/flpy/pandas/ST/@X}{0}
\pgfkeyssetvalue{/flpy/pandas/ST/time}{1.1366146802902222}
\pgfkeyssetvalue{/flpy/pandas/SBFL/time}{3809.7575186888375}
\pgfkeyssetvalue{/flpy/pandas/MBFL/time}{36560.88698505031}
\pgfkeyssetvalue{/flpy/cookiecutter/Tarantula/@X}{2}
\pgfkeyssetvalue{/flpy/cookiecutter/Ochiai/@X}{2}
\pgfkeyssetvalue{/flpy/cookiecutter/DStar/@X}{2}
\pgfkeyssetvalue{/flpy/cookiecutter/Metallaxis/@X}{0}
\pgfkeyssetvalue{/flpy/cookiecutter/Muse/@X}{0}
\pgfkeyssetvalue{/flpy/cookiecutter/PS/@X}{0}
\pgfkeyssetvalue{/flpy/cookiecutter/PS/time}{13.366540551185608}
\pgfkeyssetvalue{/flpy/cookiecutter/ST/@X}{0}
\pgfkeyssetvalue{/flpy/cookiecutter/ST/time}{1.1999731659889221}
\pgfkeyssetvalue{/flpy/cookiecutter/SBFL/time}{8.688710868358612}
\pgfkeyssetvalue{/flpy/cookiecutter/MBFL/time}{51.46605199575424}
\pgfkeyssetvalue{/flpy/fastapi/Tarantula/@X}{5}
\pgfkeyssetvalue{/flpy/fastapi/Ochiai/@X}{5}
\pgfkeyssetvalue{/flpy/fastapi/DStar/@X}{5}
\pgfkeyssetvalue{/flpy/fastapi/Metallaxis/@X}{3}
\pgfkeyssetvalue{/flpy/fastapi/Muse/@X}{3}
\pgfkeyssetvalue{/flpy/fastapi/PS/@X}{1}
\pgfkeyssetvalue{/flpy/fastapi/PS/time}{744.9599631382869}
\pgfkeyssetvalue{/flpy/fastapi/ST/@X}{1}
\pgfkeyssetvalue{/flpy/fastapi/ST/time}{0.9424397211808425}
\pgfkeyssetvalue{/flpy/fastapi/SBFL/time}{6.926379423875075}
\pgfkeyssetvalue{/flpy/fastapi/MBFL/time}{591.6665234932533}
\pgfkeyssetvalue{/flpy/keras/Tarantula/@X}{7}
\pgfkeyssetvalue{/flpy/keras/Ochiai/@X}{7}
\pgfkeyssetvalue{/flpy/keras/DStar/@X}{6}
\pgfkeyssetvalue{/flpy/keras/Metallaxis/@X}{6}
\pgfkeyssetvalue{/flpy/keras/Muse/@X}{4}
\pgfkeyssetvalue{/flpy/keras/PS/@X}{0}
\pgfkeyssetvalue{/flpy/keras/PS/time}{2977.0495265987183}
\pgfkeyssetvalue{/flpy/keras/ST/@X}{0}
\pgfkeyssetvalue{/flpy/keras/ST/time}{3.710358738899231}
\pgfkeyssetvalue{/flpy/keras/SBFL/time}{196.40626466274261}
\pgfkeyssetvalue{/flpy/keras/MBFL/time}{31330.43772265646}
\pgfkeyssetvalue{/flpy/youtube_dl/Tarantula/@X}{6}
\pgfkeyssetvalue{/flpy/youtube_dl/Ochiai/@X}{6}
\pgfkeyssetvalue{/flpy/youtube_dl/DStar/@X}{6}
\pgfkeyssetvalue{/flpy/youtube_dl/Metallaxis/@X}{7}
\pgfkeyssetvalue{/flpy/youtube_dl/Muse/@X}{8}
\pgfkeyssetvalue{/flpy/youtube_dl/PS/@X}{1}
\pgfkeyssetvalue{/flpy/youtube_dl/PS/time}{1257.269953161478}
\pgfkeyssetvalue{/flpy/youtube_dl/ST/@X}{1}
\pgfkeyssetvalue{/flpy/youtube_dl/ST/time}{3.740866184234619}
\pgfkeyssetvalue{/flpy/youtube_dl/SBFL/time}{54.46408800780773}
\pgfkeyssetvalue{/flpy/youtube_dl/MBFL/time}{6767.083820268512}
\pgfkeyssetvalue{/flpy/sanic/Tarantula/@X}{1}
\pgfkeyssetvalue{/flpy/sanic/Ochiai/@X}{1}
\pgfkeyssetvalue{/flpy/sanic/DStar/@X}{1}
\pgfkeyssetvalue{/flpy/sanic/Metallaxis/@X}{0}
\pgfkeyssetvalue{/flpy/sanic/Muse/@X}{0}
\pgfkeyssetvalue{/flpy/sanic/PS/@X}{0}
\pgfkeyssetvalue{/flpy/sanic/PS/time}{365.429003238678}
\pgfkeyssetvalue{/flpy/sanic/ST/@X}{0}
\pgfkeyssetvalue{/flpy/sanic/ST/time}{0.3602904478708903}
\pgfkeyssetvalue{/flpy/sanic/SBFL/time}{208.65312417348227}
\pgfkeyssetvalue{/flpy/sanic/MBFL/time}{11772.331229448318}
\pgfkeyssetvalue{/flpy/thefuck/Tarantula/@X}{15}
\pgfkeyssetvalue{/flpy/thefuck/Ochiai/@X}{15}
\pgfkeyssetvalue{/flpy/thefuck/DStar/@X}{15}
\pgfkeyssetvalue{/flpy/thefuck/Metallaxis/@X}{7}
\pgfkeyssetvalue{/flpy/thefuck/Muse/@X}{7}
\pgfkeyssetvalue{/flpy/thefuck/PS/@X}{0}
\pgfkeyssetvalue{/flpy/thefuck/PS/time}{48.830794125795364}
\pgfkeyssetvalue{/flpy/thefuck/ST/@X}{1}
\pgfkeyssetvalue{/flpy/thefuck/ST/time}{0.9525567293167114}
\pgfkeyssetvalue{/flpy/thefuck/SBFL/time}{5.885042876005173}
\pgfkeyssetvalue{/flpy/thefuck/MBFL/time}{72.83934625983238}
\pgfkeyssetvalue{/flpy/tqdm/Tarantula/@X}{4}
\pgfkeyssetvalue{/flpy/tqdm/Ochiai/@X}{4}
\pgfkeyssetvalue{/flpy/tqdm/DStar/@X}{4}
\pgfkeyssetvalue{/flpy/tqdm/Metallaxis/@X}{1}
\pgfkeyssetvalue{/flpy/tqdm/Muse/@X}{0}
\pgfkeyssetvalue{/flpy/tqdm/PS/@X}{0}
\pgfkeyssetvalue{/flpy/tqdm/PS/time}{191.52308593477522}
\pgfkeyssetvalue{/flpy/tqdm/ST/@X}{2}
\pgfkeyssetvalue{/flpy/tqdm/ST/time}{1.3300951208387102}
\pgfkeyssetvalue{/flpy/tqdm/SBFL/time}{41.834107228687834}
\pgfkeyssetvalue{/flpy/tqdm/MBFL/time}{7153.891149725233}
\pgfkeyssetvalue{/flpy/luigi/Tarantula/@X}{5}
\pgfkeyssetvalue{/flpy/luigi/Ochiai/@X}{5}
\pgfkeyssetvalue{/flpy/luigi/DStar/@X}{5}
\pgfkeyssetvalue{/flpy/luigi/Metallaxis/@X}{7}
\pgfkeyssetvalue{/flpy/luigi/Muse/@X}{4}
\pgfkeyssetvalue{/flpy/luigi/PS/@X}{1}
\pgfkeyssetvalue{/flpy/luigi/PS/time}{1486.3630113968482}
\pgfkeyssetvalue{/flpy/luigi/ST/@X}{2}
\pgfkeyssetvalue{/flpy/luigi/ST/time}{0.6259081730475793}
\pgfkeyssetvalue{/flpy/luigi/SBFL/time}{22.270127296447754}
\pgfkeyssetvalue{/flpy/luigi/MBFL/time}{14188.034653113438}
\pgfkeyssetvalue{/flpy/spacy/Tarantula/@X}{3}
\pgfkeyssetvalue{/flpy/spacy/Ochiai/@X}{3}
\pgfkeyssetvalue{/flpy/spacy/DStar/@X}{3}
\pgfkeyssetvalue{/flpy/spacy/Metallaxis/@X}{1}
\pgfkeyssetvalue{/flpy/spacy/Muse/@X}{2}
\pgfkeyssetvalue{/flpy/spacy/PS/@X}{1}
\pgfkeyssetvalue{/flpy/spacy/PS/time}{13915.530956427256}
\pgfkeyssetvalue{/flpy/spacy/ST/@X}{0}
\pgfkeyssetvalue{/flpy/spacy/ST/time}{0.35520780086517334}
\pgfkeyssetvalue{/flpy/spacy/SBFL/time}{60.325461665789284}
\pgfkeyssetvalue{/flpy/spacy/MBFL/time}{4920.254861911138}
\pgfkeyssetvalue{/flpy/tornado/Tarantula/@X}{2}
\pgfkeyssetvalue{/flpy/tornado/Ochiai/@X}{2}
\pgfkeyssetvalue{/flpy/tornado/DStar/@X}{2}
\pgfkeyssetvalue{/flpy/tornado/Metallaxis/@X}{1}
\pgfkeyssetvalue{/flpy/tornado/Muse/@X}{1}
\pgfkeyssetvalue{/flpy/tornado/PS/@X}{0}
\pgfkeyssetvalue{/flpy/tornado/PS/time}{1444.9314478039742}
\pgfkeyssetvalue{/flpy/tornado/ST/@X}{0}
\pgfkeyssetvalue{/flpy/tornado/ST/time}{0.7016530632972717}
\pgfkeyssetvalue{/flpy/tornado/SBFL/time}{975.7906938791275}
\pgfkeyssetvalue{/flpy/tornado/MBFL/time}{28013.037698805332}
\pgfkeyssetvalue{/flpy/black/Tarantula/@X}{4}
\pgfkeyssetvalue{/flpy/black/Ochiai/@X}{4}
\pgfkeyssetvalue{/flpy/black/DStar/@X}{4}
\pgfkeyssetvalue{/flpy/black/Metallaxis/@X}{5}
\pgfkeyssetvalue{/flpy/black/Muse/@X}{1}
\pgfkeyssetvalue{/flpy/black/PS/@X}{2}
\pgfkeyssetvalue{/flpy/black/PS/time}{45148.89834061036}
\pgfkeyssetvalue{/flpy/black/ST/@X}{1}
\pgfkeyssetvalue{/flpy/black/ST/time}{0.7087001250340388}
\pgfkeyssetvalue{/flpy/black/SBFL/time}{61.91867428559524}
\pgfkeyssetvalue{/flpy/black/MBFL/time}{28935.937529087067}
\pgfkeyssetvalue{/flpy/all/Tarantula/@X}{58}
\pgfkeyssetvalue{/flpy/all/Ochiai/@X}{58}
\pgfkeyssetvalue{/flpy/all/DStar/@X}{57}
\pgfkeyssetvalue{/flpy/all/Metallaxis/@X}{40}
\pgfkeyssetvalue{/flpy/all/Muse/@X}{34}
\pgfkeyssetvalue{/flpy/all/PS/@X}{9}
\pgfkeyssetvalue{/flpy/all/PS/time}{9751.161942639174}
\pgfkeyssetvalue{/flpy/all/ST/@X}{8}
\pgfkeyssetvalue{/flpy/all/ST/time}{1.603535818170618}
\pgfkeyssetvalue{/flpy/all/SBFL/time}{589.0094189343629}
\pgfkeyssetvalue{/flpy/all/MBFL/time}{15774.390040752623}

\newcommand{\reducedstrut}{%
  \vrule width 0pt height .9\ht\strutbox depth .9\dp\strutbox\relax%
}

\newcommand{\best}[2][bcol!50]{%
  \begingroup
  \setlength{\fboxsep}{0.5pt}%
  \colorbox{#1}{\reducedstrut#2\/}%
  \endgroup
}

\DeclareDocumentCommand{\n}{t. t: o m o O{} t| t!}{%
  \begingroup%
  \pgfkeys{/pgf/fpu=true}%
  \IfBooleanTF{#2}{%
    \pgfkeyssetvalue{/tmp/value}{#4}%
    \pgfkeyssetvalue{/tmp/found}{found}%
  }{%
    \pgfkeysifdefined{\keyfamily#4}{%
      \pgfkeyssetvalue{/tmp/value}{\pgfkeysvalueof{\keyfamily#4}}%
      \pgfkeyssetvalue{/tmp/found}{found}%
    }{}%
  }%
  \pgfkeysifdefined{/tmp/found}{%
    \IfNoValueF{#5}{%
      \pgfkeyssetvalue{/tmp/multiplier}{#5}%
      \pgfmathparse{\pgfkeysvalueof{/tmp/multiplier} * \pgfkeysvalueof{/tmp/value}}%
      \pgfkeyslet{/tmp/value}\pgfmathresult%
    }%
    \IfBooleanTF{#1}{%
      \IfBooleanTF{#8}{%
        \spellout{\pgfkeysvalueof{/tmp/value}}}{%
        \pgfkeysvalueof{/tmp/value}}%
    }{%
      \IfNoValueTF{#3}{%
        \pgfmathprintnumber%
        [set thousands separator={\,},int detect,#6]%
        {\pgfkeysvalueof{/tmp/value}}%
      }{%
        {\pgfmathprintnumber%
          [precision=#3,fixed,zerofill,set thousands separator={\,},#6]%
          {\pgfkeysvalueof{/tmp/value}}}%
      }%
    }%
    \IfBooleanT{#7}{{\smaller[1.2]\%}}%
  }{%
    \NAN%
  }%
  \pgfkeys{/pgf/fpu=false}%
  \endgroup%
}

\begin{document}

\title{\fp: A Fault Localization Tool for Python
\thanks{Work partially supported by SNF grant 200021-182060 (Hi-Fi).}
}

\author{\IEEEauthorblockN{Mohammad Rezaalipour and Carlo A. Furia}\\
	\IEEEauthorblockA{\textit{Software Institute} -- 
		\textit{USI Universit\`{a} della Svizzera italiana}\\
		Lugano, Switzerland\\
		\{rezaam, furiac\}@usi.ch}
}

\maketitle

\begin{abstract}
This paper presents \fp, a fault localization tool for Python programs.
\fp supports seven well-known fault localization techniques
in four families:
spectrum-based, mutation-based, predicate switching, and stack trace
fault localization.
It is implemented as plugin of the popular Pytest testing framework,
but also works with tests written for Unittest and Hypothesis
(two other popular testing frameworks).
The paper showcases how to use \fp on two illustrative examples,
and then discusses its main features and capabilities from a user's perspective.
To demonstrate that \fp is applicable to analyze Python projects
of realistic size, the paper also summarizes the results of an extensive
experimental evaluation that applied \fp to 135 real-world bugs
from the \bip curated collection.
To our knowledge, \fp is the first open-source fault localization tool for Python
that supports multiple fault localization families.
\end{abstract}


\section{Introduction}
\label{sec:introduction}

Starting from around the 1990s~\cite{FL-survey},
there has been a growing interest in automated \emph{fault localization}
techniques for programs,
which spurred the development of increasingly sophisticated
and effective techniques.
Nowadays, fault localization techniques are widely used
both on their own, and as components of more complex
(dynamic) program analyses---for example,
as ingredients of automated program repair.

Like with every program analysis technique,
practical adoption of fault localization
critically requires
that reusable, flexible \emph{tool} implementations
are available,
so that trying out new research ideas and applications
does not require to re-implement from scratch techniques
that are already known to work.
Somewhat unexpectedly, however,
the lion's share of work in fault localization
targets languages like C and Java
(as we discuss in \autoref{sec:related-work});
in contrast, there is comparatively little work%
---and, most important, barely any available tools---%
for the popular Python programming language.

To address this deficiency,
this paper describes \fp (read: ``foh pie''):
a fault localization tool for Python.
To our knowledge,
\fp is the only available Python fault localization
tool that supports multiple fault localization families
(spectrum based, mutation based, stack-trace based, and predicate switching).
The immediate motivation for implementing \fp
was to carry out a large scale empirical study
of fault localization in Python,
which we describe in a separate paper~\cite{Rezaalipour:2023}.
Nevertheless, we designed and implemented \fp
with the broader goal
of making it a flexible, reusable stand-alone tool
for all applications of fault localization in Python.

The current paper presents, focusing on the user's perspective,
the tool \fp,
some of its concrete usage scenarios (\autoref{sec:using}),
and its main features and implementation (\autoref{sec:architecture}).
\fp supports seven fault localization techniques,
and two localization granularities (statement and function);
it can use tests written for the most popular Python unit testing frameworks
such as Pytest and Unittest;
it can be extended with support for new techniques.
To demonstrate that \fp is applicable to real-world
projects, we also summarize some of the results
of our recent empirical study~\cite{Rezaalipour:2023}
that used it.
As we detail in \autoref{sec:availability},
\fp is available as open source.
A short demo of \fp is available at \url{https://youtu.be/O4T7w-U8rZE}.

\section{Using \fp}
\label{sec:using}

This section overviews using \fp on two simple examples,
from the perspective of Moe---a nondescript user.

\subsection{Spectrum-based and Mutation-based Fault Localization}
To practice programming in Python,
Moe has implemented function \Py{equilateral_area} in \autoref{lst:motivational1}.
The function takes as input the length \Py{side} of an equilateral
triangle's side, and returns its area computed using the formula $\Py{side}^2 \times \sqrt{3}/4$.
Unfortunately, Moe inadvertently introduced a bug\footnote{In the paper, we use the terms ``fault'' and ``bug'' as synonyms.}
on line~\ref{l:bug:1}, which sums variables \Py{const} and \Py{term}
instead of \emph{multiplying} them.
Fortunately, the bug does not go unnoticed thanks to the
tests that Moe also wrote (see \autoref{lst:motivational1:tests});
in particular, the assertion in test \Py{test_ea_fail} fails,
indicating that \Py{equilateral_area} does not work as intended.

\begin{lstlisting}[float=!tb,caption=
Python function \Py{equilateral_area} computes the area of
an equilateral triangle given its \Py{side} length; this implementation has a bug at line~\ref{l:bug:1}.,
label={lst:motivational1}]
def equilateral_area(side):
    const = math.sqrt(3) / 4
    if side == 1:  (*\label{l:if}*)
        return const (*\label{l:return}*)
    term = math.pow(side, 2)   (*\label{l:term}*)
    area = const (*\textcolor{red}{+}*) term        (*\textcolor{red}{\# bug} \label{l:bug:1}*)
    return area     (*\label{l:return-area}*)
\end{lstlisting}

\begin{lstlisting}[float=!tb,caption=
Tests for function \Py{equilateral_area} in \autoref{lst:motivational1}. Library function \Py{pytest.approx} checks equality of floating points within some tolerance.,label={lst:motivational1:tests}]
def test_ea_fail():
    area = equilateral_area(side=3)
    assert area == pytest.approx(9 * math.sqrt(3) / 4)

def test_ea_pass():
    area = equilateral_area(side=1)
    assert area == pytest.approx(math.sqrt(3) / 4)
\end{lstlisting}

To help him debug \Py{equilateral_area},
Moe runs our fault localization tool \fp.
All fault localization techniques implemented by \fp
are \emph{dynamic} (i.e., based on tests);
therefore, Moe points \fp to the location
of \Py{equilateral_area}'s implementation,
as well as of its \emph{tests}.
By default,
\fp performs spectrum-based fault localization
(SBFL)---a family of widely used fault localization techniques
based on the idea of comparing program traces (``spectra'')
of passing and failing runs of a program.
\fp currently supports three techniques
(DStar~\cite{Wong:2014}, Ochiai~\cite{Abreu:2007}, Tarantula~\cite{Jones:2005})
that belong to the SBFL family;
since they only differ in the formula used to aggregate the
information about traces, \fp reports the output for all SBFL techniques
with a single analysis run.

SBFL runs quite fast, taking only 0.3 seconds on this example.
The output, like for every fault localization technique,
is a list of program locations (identified by line numbers)
ranked by their \emph{suspiciousness score};
the absolute value of the suspiciousness score does not matter,
what matters is the \emph{rank} of a location:
the higher its rank, the more likely the location
is implicated with the failure triggered by the tests.
As shown in \autoref{tab:example-results},
all three SBFL techniques correctly assign the top rank
to the fault location (line~\ref{l:bug:1} in \autoref{lst:motivational1});
however, they also assign the top rank to some nearby locations
(lines \ref{l:term} and \ref{l:return-area})
which tie the faulty location's suspiciousness score.
This example highlights a fundamental limitation of SBFL techniques:
since they compare traces in different executions,
they cannot distinguish between locations
that are in the same basic block
(a portion of code without branches).

\begin{table*}[!tb]
  \centering
  \setlength{\tabcolsep}{8pt}
  \begin{tabular}{c r *{7}{r}}
    \toprule
    && \multicolumn{2}{c}{MBFL} & \multicolumn{3}{c}{SBFL} &  & \\
    \cmidrule(lr){3-4} \cmidrule(lr){5-7} \cmidrule(lr){8-8} \cmidrule(lr){9-9}
    \textsc{example} & & \multicolumn{1}{c}{Metallaxis} & \multicolumn{1}{c}{Muse}
     & \multicolumn{1}{c}{DStar} & \multicolumn{1}{c}{Ochiai} & \multicolumn{1}{c}{Tarantula} & \multicolumn{1}{c}{PS} & \multicolumn{1}{c}{ST} \\
    \midrule
    \multirow{2}{*}{\autoref{lst:motivational1}} &\textsc{time} [seconds] & \multicolumn{2}{c}{15.9} & \multicolumn{3}{c}{0.3} & 1.2 & 0.2 \\
    \cmidrule(lr){3-4} \cmidrule(lr){5-7} \cmidrule(lr){8-8} \cmidrule(lr){9-9}
    &\textsc{top-rank locations} & \ref{l:if} \ref{l:term} \best{\ref{l:bug:1}} & \best{\ref{l:bug:1}} & \ref{l:term}, \best{\ref{l:bug:1}}, \ref{l:return-area} & \ref{l:term}, \best{\ref{l:bug:1}}, \ref{l:return-area} & \ref{l:term}, \best{\ref{l:bug:1}}, \ref{l:return-area} & -- & -- \\
    \midrule
    \multirow{2}{*}{\autoref{lst:motivational2}} &\textsc{time} [seconds] & \multicolumn{2}{c}{18.4} & \multicolumn{3}{c}{0.1} & 0.2 & 0.1 \\
    \cmidrule(lr){3-4} \cmidrule(lr){5-7} \cmidrule(lr){8-8} \cmidrule(lr){9-9}
    &\textsc{top-rank locations} & \best{\ref{l:bug:2}}, \ref{l:h:3}, \ref{l:ia:exp} & \best{\ref{l:bug:2}}, \ref{l:h:3}, \ref{l:ia:exp} & \ref{l:h:1}, \best{\ref{l:bug:2}}, \ref{l:h:3}, \ref{l:ia:exp} & \ref{l:h:1}, \best{\ref{l:bug:2}}, \ref{l:h:3}, \ref{l:ia:exp} & \ref{l:h:1}, \best{\ref{l:bug:2}}, \ref{l:h:3}, \ref{l:ia:exp} & -- & \ref{l:h:1}, \best{\ref{l:bug:2}}, \ref{l:h:3} \\
    \bottomrule
  \end{tabular}
  \caption{A summary of running \fp on the two examples in \autoref{lst:motivational1} (\Py{equilateral_area}) and \autoref{lst:motivational2} (\Py{isosceles_area}). For each fault localization technique (grouped by family), the table reports the running \textsc{time} of \fp in seconds, and the program locations (line numbers) with the highest suspiciousness (\textsc{top-rank}). A colored background \best{highlights} the actual location of the bug in each example. Since the running time of all techniques in a family is the same, it is only reported once per family.}
  \label{tab:example-results}
\end{table*}

Mutation-based fault localization (MBFL) techniques
use a different approach,
which is capable of distinguishing between
locations in the same basic block.
As the name suggests, MBFL techniques are based
on mutation testing:
given a program to analyze,
they generate many different \emph{mutants}%
---syntactic mutations obtained
by systematically applying a number of mutation operators.
The intuition is that if mutating the code at a certain program
location changes the program behavior
(a test passes on the original program and fails on the mutant, or vice versa),
then the program location is likely to be implicated with the fault.

To run MBFL with \fp,
Moe simply adds the option \verb+--family mbfl+.
\fp currently supports two techniques
(Metallaxis~\cite{Papadakis:2015} and Muse~\cite{Moon:2014})
that belong to the MBFL family;
just like for SBFL techniques,
a single analysis run of \fp computes the output of both MBFL techniques.
MBFL is notoriously time consuming;
in fact, it takes 15.9 seconds on \autoref{lst:motivational1}'s example
(over 50 times longer than SBFL).
As shown in \autoref{tab:example-results},
the two MBFL techniques
achieve quite different results despite using the same 32 mutants for analysis:
Muse is very accurate, as it singles out line~\ref{l:bug:1}
as the most suspicious location;
Metallaxis also ranks it at the top,
but together with two other locations that are not responsible for the fault.

\subsection{Stack Trace and Predicate Switching Fault Localization}

\fp supports two other fault-localization families:
stack-trace (ST~\cite{Zou:2021}) fault localization
and predicate switching (PS~\cite{Zhang:2006}).\footnote{%
Unlike SBFL and MBFL, there is only one implementation of ST and one of PS%
---hence, ST and PS denote both families and individual techniques.
}
Moe tries them out on \Py{equilateral_area}
but the results are disappointing:
both techniques return the empty list of locations,
meaning that they could not gather any evidence of suspiciousness.
The reason for ST's failure in this case is quite obvious:
ST analyzes the stack trace dumped after a program \emph{crash}
(usually, an uncaught exception);
since all tests in \autoref{lst:motivational1:tests} terminate without
crashing, ST is completely ineffective on this example.

In order to try an example where ST may stand a chance,
Moe considers another little Python program he wrote:
function \Py{isosceles_area} 
returns the area of an isosceles triangle
computed as $\Py{base}/2 \times \sqrt{\Py{leg}^2 - \Py{base}^2/4}$.\footnote{
  The \emph{legs} of an isosceles triangle are the two sides of equal length;
  the third side is called \emph{base}.
}
The implementation in \autoref{lst:motivational2}
erroneously swaps \Py{base} and \Py{leg};
thus, when executing test \Py{test_ia_crash},
expression \Py{t1 - t2} in \autoref{lst:motivational2}
evaluates to a negative number,
which crashes the program with an uncaught \Py{ValueError}
exception raised by \Py{math.sqrt}.

\begin{lstlisting}[float=!tb,caption=
Python function \Py{isosceles_area} computes the area of
an isosceles triangle given its \Py{leg} and \Py{base} lengths; this implementation has a bug at line~\ref{l:bug:2}.,
label={lst:motivational2}]
def isosceles_area(leg, base):
    def height():(*\label{l:h:1}*)
        t1, t2 = math.pow((*\textcolor{red}{base}*), 2), math.pow((*\textcolor{red}{leg}*), 2) / 4    (*\textcolor{red}{\# bug} \label{l:bug:2}*)
        return math.sqrt(t1 - t2)(*\label{l:h:3}*)

    area = 0.5 * base * height()(*\label{l:ia:exp}*)
    return area
\end{lstlisting}

\begin{lstlisting}[float=!tb,caption=
Tests for function \Py{isosceles_area} in \autoref{lst:motivational2}. Library function \Py{pytest.approx} checks equality of floating points within some tolerance.,label={lst:motivational2:tests}]
def test_ia_crash():
    area = isosceles_area(leg=9, base=4)
    assert area == pytest.approx(2 * math.sqrt(77))  (*\label{l:assertion}*)

def test_ia_pass():
    area = isosceles_area(leg=4, base=4)
    assert area == pytest.approx(2 * math.sqrt(12))
\end{lstlisting}

Executing \fp with option \verb+--family st+
on \autoref{lst:motivational2}'s example
terminates quickly (around 0.1 seconds)
and ranks the three locations \ref{l:h:1}, \ref{l:bug:2}, \ref{l:h:3}
as top suspiciousness.
Even this simple example showcases
ST's key features:
first, it is usually very fast,
since it does not have to collect any information
other than the stack trace of crashing tests.
Second, it can be quite effective
with crashing bugs (after all, the fault location \ref{l:bug:2}
is ranked at the top),
but fundamentally operates at the level of whole \emph{functions}:
a stack trace reports the list of functions that were active
when the program crashed;
hence, ST fault localization cannot distinguish
between the suspiciousness of locations that belong to the same function
(\Py{height} in the example, which consists of three lines).
Still, on this example,
ST is a bit more accurate than SBFL
(which also ranks line~\ref{l:ia:exp} in the top position),
and arguably somewhat better than MBFL
(which also reports three locations at the top rank,
but one of them is line~\ref{l:ia:exp}, which is the call location of \Py{height},
and hence not really responsible for the fault).
The running time of ST and SBFL is practically indistinguishable
on this simple example; in general, however, SBFL takes more time than ST
because the latter only runs failing tests
and does not require any \emph{tracing} when executing the tests.
As usual, MBFL takes considerably longer (18.4 seconds)
to generate several mutants (48 mutants)
and to execute all tests on each mutant.

As a last experiment of \fp's capabilities,
Moe runs PS fault localization on \Py{isosceles_area}.
Just like on \Py{equilateral_area},
PS fails to localize the bug and returns an empty list of locations.
Once again, the program features explain why these examples
are a poor match for PS's capabilities.
As the name suggests,
PS is based on the idea of
\emph{forcefully changing}
the outcome of a program conditional branch
dynamically during different test executions;
the intuition is that if switching a \emph{predicate} (branch condition)
turns a failing test into a passing one,
then the predicate may be responsible for the fault.
Clearly, if a program has no conditionals
(like \Py{isosceles_area}),
or its conditionals are unrelated to the locations of failure
(like \Py{equilateral_area}),
PS is unlikely to be of any help to locate the bug.

\iflong
In all, this section's simple examples
gave a concrete idea of \fp's capabilities,
showcasing the variety of fault localization techniques
that it supports and how they can be applied.
\fi

\section{\fp's Architecture and Implementation}
\label{sec:architecture}

\fp is an automated fault localization tool for Python.
The current version of \fp supports
seven fault localization techniques in four families:
the spectrum-based (SBFL) techniques
DStar~\cite{Wong:2014}, Ochiai~\cite{Abreu:2007}, and Tarantula~\cite{Jones:2005};
the mutation-based (MBFL) techniques
\muse~\cite{Moon:2014} and Metallaxis~\cite{Papadakis:2015};
and the predicate switching~(PS)~\cite{Zhang:2006}
and stack trace~(ST)~\cite{Zou:2021} fault localization families/techniques.

\fp can perform fault localization with two granularities:
statement-level and function-level.
That is, the granularity determines what program \emph{entities}
are localized:
the locations of individual statements in the source code, or
the functions that compose the programs.

\fp is a command-line tool,
implemented as a plugin of the popular Pytest testing framework.
As essential input,
\fp takes the location of
the source code of a Python project where to perform fault localization,
as well as the location of a test suite.
\fp accepts tests in the formats of Pytest,
as well as Unittest (another widely used Python testing framework)
and Hypothesis (a property-based Python testing tool, which supports the definition of parametric tests).
As output, \fp returns a CSV file
listing program entities ranked by their suspiciousness score;
the higher an entity's suspiciousness score,
the more likely the entity is the location of the fault.

\subsection{Features and Options}
\label{sec:options}

The only mandatory command line argument to use \fp
is \Bash{--src PACKAGE},
which runs SBFL at the statement granularity
on the Python package in directory \Bash{PACKAGE},
using any tests discovered by Pytest within the project's source files.

Flags \Bash{--family}
and \Bash{--granularity}
respectively select the fault localization family
(SBFL, MBFL, ST, and PS)
and the granularity
(statement and function)
at which to perform fault localization.
As mentioned in \autoref{sec:using},
\fp simultaneously runs all techniques
that belong to the selected family,
since it is able to reuse the output of the same underlying analysis.

Using Pytest's command line options,
users can select specific tests to be used by \fp.
For example, you can run a test selection algorithm to
identify a subset of the tests,
and then feed its output to \fp.
The command line option \Bash{--failing-list}
explicitly asks \fp to only use the given list of \emph{failing} tests.
Normally, \fp runs all available tests, and figures out which are passing
and which are failing.
However, a technique like ST only needs to run failing tests;
thus, if those are given to \fp explicitly,
ST can run much faster by simply ignoring all passing tests.
Another scenario where selecting failing tests is useful is
whenever a test suite includes multiple failing tests
that trigger \emph{different} bugs;
localizing one bug at a time is likely to increase
the effectiveness of fault localization techniques%
---whose heuristics usually assume that all failures
refer to the same fault.

\begin{figure}[!tb]
\centering
\begin{tikzpicture}[
  stage/.style args={#1/#2}{rectangle,fill=ecol!40,
    font=\footnotesize\bfseries,text=black, draw=none,
    label={[font=\scriptsize\sffamily,label distance=0pt]below:#1},
    label={[font=\scriptsize,label distance=-1pt]above:#2}},
  technique/.style={rectangle,fill=acol,
    font=\footnotesize\bfseries,text=white, draw=none},
  data/.style args={#1/#2}{draw=none,fill=ccol,font=\footnotesize,text=white,
    label={[font=\scriptsize\ttfamily,label distance=1pt]below:#1},
    label={[font=\scriptsize\ttfamily,label distance=-1pt,xshift=2mm]above:#2}},
  align=center,
  ]
  \matrix (allnodes) [row sep=5mm, column sep=3.5mm, ampersand replacement=\&]
  {
    \&
    \&
    \&
    \node [stage={ps\_inst/\faRandom}] (ps-inst) {instrument\\predicates};
    \&
    \node [technique] (ps) {PS};
    \\
    \&
    \&
    \node [stage={Pytest/{\color{green}\faCheck} {\color{red}\faClose} \faBug}] (run-tests) {run tests};
    \&
    \&
    \node [technique] (st) {ST};
    \\
    \node [data={py \faFlask/}] (tests) {tests};
    \&
    \&
    \&
    \node [stage={Coverage.py/\faCodeFork}] (coverage) {calculate\\coverage};
    \&
    \node [technique] (sbfl) {SBFL};
    \&
    \&
    \node [data={csv \faDatabase/\faSortAmountDesc}] (csv) {ranked\\suspicious\\locations};
    \\
    \node [data={py \faFileTextO/}] (program) {program};
    \&
    \&
    \&
    \node [stage={Cosmic Ray/\twemoji{alien monster}}] (mutants) {generate\\mutants};
    \&
    \node [technique] (mbfl) {MBFL};
    \&
    \&
    \&
    \\
  };

  \node [fit=(run-tests)(ps-inst)(coverage)(mutants)(sbfl)(mbfl)(st)(ps),
  draw=acol,ultra thick, rounded corners,inner xsep=3mm,
  inner ysep=4mm,
  label={[acol]north:\fp}] (fp) {};

  \coordinate (mid-ps) at ($(tests.center)!0.72!(ps.center)$);
  \coordinate (mid-mbfl) at ($(tests.center)!0.65!(mbfl.center)$);
  \coordinate (output) at ($(sbfl.center)!0.33!(csv.center)$);
  \coordinate (tests-input) at ($(tests.center)!0.31!(coverage.center)$);

  \coordinate (tests-out) at (tests.center -| fp.west);
  
  \coordinate (program-out) at (program.center -| fp.west);

  \coordinate (overlap-h) at ($(fp.west)!0.35!(run-tests.west)$);
  \coordinate (overlap) at (tests-out.center -| overlap-h);

  \coordinate (csv-out) at (csv.center -| fp.east);

  \draw (ps) -| (output);
  \draw (st) -| (output);
  \draw (sbfl) -- (csv-out);
  \draw (mbfl) -| (output);
  \draw (tests-out) -- (tests-input);

  \begin{scope}[->]

    \begin{scope}[very thick,acol]
      \draw (tests) -- (tests-out);
      \draw (program) -- (program-out);
      \draw (csv-out) -- (csv);
    \end{scope}
    
    \draw (tests-input) |- (run-tests.west);
    \draw (program) -- (mutants);
    \draw (program-out) -| ($(overlap)+(0,-1.5pt)$) arc[start angle=270, end angle=90, radius=1.5pt] |- (ps-inst.west);

    \draw (run-tests.east) -- (coverage);
    \draw (run-tests) -- (st);
    \draw (coverage) -- (sbfl);
    \draw (mutants) -- (mbfl);
    \draw (ps-inst) -- (ps);
    \draw (tests-input) |- (mid-mbfl) -| (mbfl.north);
    \draw (tests-input) |- (mid-ps) -| (ps.south);
  \end{scope}
\end{tikzpicture}
\caption{An overview of \fp's architecture.}
\label{fig:workflow}
\end{figure}
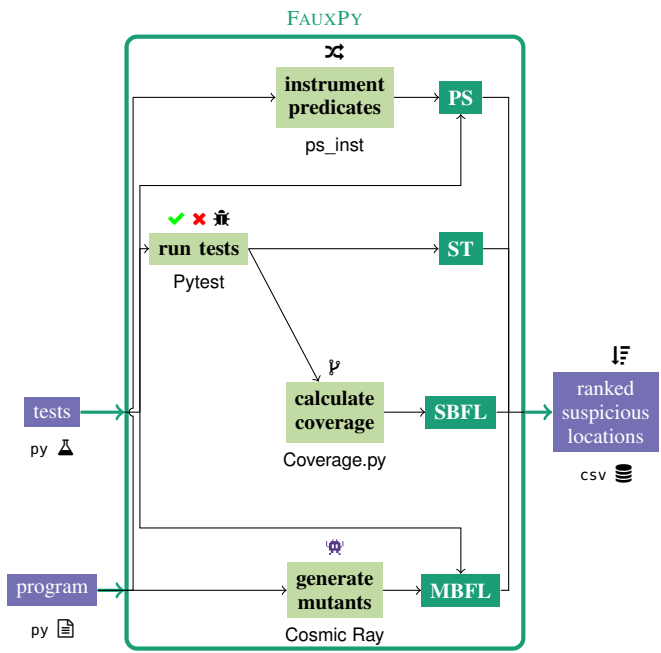

\subsection{Implementation}
\label{sec:implementation}

\autoref{fig:workflow}
overviews the workflow of \fp.
The first step of \fp's dynamic analysis
is always running the available tests.
Then, different components collect different kind of information
required by the selected fault localization technique.

SBFL techniques rely on coverage information;
to this end,
\fp runs \project{Coverage.py}~\cite{Batchelder:2023},
a popular coverage library for Python.
MBFL techniques
generate several mutants of the input program;
to this end,
\fp uses state-of-the-art mutation framework \project{Cosmic Ray}~\cite{CosmicRay:2023}.
By default, \fp applies the framework's default mutation operators,
but users can also provide other custom operators.
\fp includes a module \project{ps\_inst}
that we developed to generate the kind of instrumentation
needed by PS fault localization;
our implementation is based on Python's \Py{ast} library.
\fp's support of ST fault localization 
parses the dumped output of all crashing tests,
reconstructs the stack trace,
and then locates the corresponding functions in the program's source code.

\fp outputs the results of its fault localization
analysis in CSV format,
encoding a ranked list of program entities
and their suspiciousness scores.
For performance reason, \fp stores all the intermediate results
(the outcome of running the various analyses and tools)
of its analysis in an SQLite database file.
This SQLite database remains available to the user
after \fp terminates executing,
which can be useful both for debugging
and to perform additional analyses on
the large amount of data collected by \fp's dynamic analysis.

\begin{table*}[!tb]
\small
\setlength{\tabcolsep}{2.5pt}
\centering
\caption{Overview of \fp's experimental evaluation on an ample selection of bugs from \bip~\cite{Widyasari:2020}. For each \textsc{project},
  the table shows its size in \textsc{kLOC},
  number of \textsc{tests}, number of \textsc{faults} analyzed with \fp,
  and how many of them each technique correctly localized
  within the top-5 positions ($@5$ \textsc{count}),
  and the \textsc{average time} per fault taken by \fp (by fault localization family, since all techniques in a family take the same time).}
    \begin{tabular}{lrrrrrrrrrrrrrr}
    \toprule

    \multicolumn{1}{c}{\multirow{3}{*}{\textsc{project}}} & 
    \multicolumn{1}{c}{\multirow{3}{*}{\textsc{kLOC}}} & 
    \multicolumn{1}{c}{\multirow{3}{*}{\textsc{tests}}} & 
    \multicolumn{1}{c}{\multirow{3}{*}{\textsc{faults}}} & 
    \multicolumn{7}{c}{$@5$ \textsc{count on project}} &
    \multicolumn{4}{c}{\textsc{average time/fault} [sec]} \\

    \cmidrule(lr){5-11} 
    \cmidrule(lr){12-15}
    
    &        
    &
    &
    &
    \multicolumn{2}{c}{MBFL} &
    \multicolumn{1}{c}{PS} &
    \multicolumn{3}{c}{SBFL} &
    \multicolumn{1}{c}{ST} &
    \multicolumn{1}{c}{{\multirow{2}{*}{MBFL}}} &
    \multicolumn{1}{c}{{\multirow{2}{*}{PS}}} &
    \multicolumn{1}{c}{{\multirow{2}{*}{SBFL}}} &
    \multicolumn{1}{c}{{\multirow{2}{*}{ST}}} \\

    \cmidrule(lr){5-6} 
    \cmidrule(lr){7-7}
    \cmidrule(lr){8-10}
    \cmidrule(lr){11-11}
    
    &        
    &
    &
    &
    \multicolumn{1}{c}{Metallaxis} &
    \multicolumn{1}{c}{Muse} &
    \multicolumn{1}{c}{PS} &
    \multicolumn{1}{c}{DStar} &
    \multicolumn{1}{c}{Ochiai} &
    \multicolumn{1}{c}{Tarantula} &
    \multicolumn{1}{c}{ST} &
    &
    &
    &
    \\
    
    \midrule
    
\project{black}          & $96.0$   & $142$    & $13$   &
\n[0]{black/Metallaxis/@X}         &
\n[0]{black/Muse/@X}   &
\n[0]{black/PS/@X} &
\n[0]{black/DStar/@X}    &
\n[0]{black/Ochiai/@X}     &
\n[0]{black/Tarantula/@X}        &
\n[0]{black/ST/@X} &
\n[0]{black/MBFL/time}   &
\n[0]{black/PS/time} &
\n[0]{black/SBFL/time}  &
\n[0]{black/ST/time} \\

\project{cookiecutter}   & $4.7$    & $300$    & $4$    &
\n[0]{cookiecutter/Metallaxis/@X}         &
\n[0]{cookiecutter/Muse/@X}   &
\n[0]{cookiecutter/PS/@X} &
\n[0]{cookiecutter/DStar/@X}    &
\n[0]{cookiecutter/Ochiai/@X}     &
\n[0]{cookiecutter/Tarantula/@X}        &
\n[0]{cookiecutter/ST/@X} &
\n[0]{cookiecutter/MBFL/time}   &
\n[0]{cookiecutter/PS/time} &
\n[0]{cookiecutter/SBFL/time}  &
\n[0]{cookiecutter/ST/time} \\

\project{fastapi}        &  $25.3$   & $842$    & $13$   &
\n[0]{fastapi/Metallaxis/@X}         &
\n[0]{fastapi/Muse/@X}   &
\n[0]{fastapi/PS/@X} &
\n[0]{fastapi/DStar/@X}    &
\n[0]{fastapi/Ochiai/@X}     &
\n[0]{fastapi/Tarantula/@X}        &
\n[0]{fastapi/ST/@X} &
\n[0]{fastapi/MBFL/time}   &
\n[0]{fastapi/PS/time} &
\n[0]{fastapi/SBFL/time}  &
\n[0]{fastapi/ST/time} \\

\project{httpie}         &
$5.6$    & $309$    & $4$    &
\n[0]{httpie/Metallaxis/@X}         &
\n[0]{httpie/Muse/@X}   &
\n[0]{httpie/PS/@X} &
\n[0]{httpie/DStar/@X}    &
\n[0]{httpie/Ochiai/@X}     &
\n[0]{httpie/Tarantula/@X}        &
\n[0]{httpie/ST/@X} &
\n[0]{httpie/MBFL/time}   &
\n[0]{httpie/PS/time} &
\n[0]{httpie/SBFL/time}  &
\n[0]{httpie/ST/time} \\

\project{keras}          &
$48.2$   & $841$    & $18$   & 
\n[0]{keras/Metallaxis/@X}         &
\n[0]{keras/Muse/@X}   &
\n[0]{keras/PS/@X} &
\n[0]{keras/DStar/@X}    &
\n[0]{keras/Ochiai/@X}     &
\n[0]{keras/Tarantula/@X}        &
\n[0]{keras/ST/@X} &
\n[0]{keras/MBFL/time}   &
\n[0]{keras/PS/time} &
\n[0]{keras/SBFL/time}  &
\n[0]{keras/ST/time} \\

\project{luigi}          &
$41.5$   & $\numprint{1718}$  & $13$   & 
\n[0]{luigi/Metallaxis/@X}         &
\n[0]{luigi/Muse/@X}   &
\n[0]{luigi/PS/@X} &
\n[0]{luigi/DStar/@X}    &
\n[0]{luigi/Ochiai/@X}     &
\n[0]{luigi/Tarantula/@X}        &
\n[0]{luigi/ST/@X} &
\n[0]{luigi/MBFL/time}   &
\n[0]{luigi/PS/time} &
\n[0]{luigi/SBFL/time}  &
\n[0]{luigi/ST/time} \\

\project{pandas}         &
$292.2$  & $\numprint{70333}$  & $18$   & 
\n[0]{pandas/Metallaxis/@X}         &
\n[0]{pandas/Muse/@X}   &
\n[0]{pandas/PS/@X} &
\n[0]{pandas/DStar/@X}    &
\n[0]{pandas/Ochiai/@X}     &
\n[0]{pandas/Tarantula/@X}        &
\n[0]{pandas/ST/@X} &
\n[0]{pandas/MBFL/time}   &
\n[0]{pandas/PS/time} &
\n[0]{pandas/SBFL/time}  &
\n[0]{pandas/ST/time} \\

\project{sanic}          &
$14.1$   & $643$    & $3$    & 
\n[0]{sanic/Metallaxis/@X}         &
\n[0]{sanic/Muse/@X}   &
\n[0]{sanic/PS/@X} &
\n[0]{sanic/DStar/@X}    &
\n[0]{sanic/Ochiai/@X}     &
\n[0]{sanic/Tarantula/@X}        &
\n[0]{sanic/ST/@X} &
\n[0]{sanic/MBFL/time}   &
\n[0]{sanic/PS/time} &
\n[0]{sanic/SBFL/time}  &
\n[0]{sanic/ST/time} \\

\project{spaCy}          &
$102.0$  & $\numprint{1732}$   & $6$    &
\n[0]{spacy/Metallaxis/@X}         &
\n[0]{spacy/Muse/@X}   &
\n[0]{spacy/PS/@X} &
\n[0]{spacy/DStar/@X}    &
\n[0]{spacy/Ochiai/@X}     &
\n[0]{spacy/Tarantula/@X}        &
\n[0]{spacy/ST/@X} &
\n[0]{spacy/MBFL/time}   &
\n[0]{spacy/PS/time} &
\n[0]{spacy/SBFL/time}  &
\n[0]{spacy/ST/time} \\

\project{thefuck}        &
$11.4$   & $\numprint{1741}$   & $16$   &
\n[0]{thefuck/Metallaxis/@X}         &
\n[0]{thefuck/Muse/@X}   &
\n[0]{thefuck/PS/@X} &
\n[0]{thefuck/DStar/@X}    &
\n[0]{thefuck/Ochiai/@X}     &
\n[0]{thefuck/Tarantula/@X}        &
\n[0]{thefuck/ST/@X} &
\n[0]{thefuck/MBFL/time}   &
\n[0]{thefuck/PS/time} &
\n[0]{thefuck/SBFL/time}  &
\n[0]{thefuck/ST/time} \\

\project{tornado}        &
$27.7$   & $\numprint{1160}$   & $4$    &
\n[0]{tornado/Metallaxis/@X}         &
\n[0]{tornado/Muse/@X}   &
\n[0]{tornado/PS/@X} &
\n[0]{tornado/DStar/@X}    &
\n[0]{tornado/Ochiai/@X}     &
\n[0]{tornado/Tarantula/@X}        &
\n[0]{tornado/ST/@X} &
\n[0]{tornado/MBFL/time}   &
\n[0]{tornado/PS/time} &
\n[0]{tornado/SBFL/time}  &
\n[0]{tornado/ST/time} \\

\project{tqdm}           &
$4.8$    & $88$     & $7$    &
\n[0]{tqdm/Metallaxis/@X}         &
\n[0]{tqdm/Muse/@X}   &
\n[0]{tqdm/PS/@X} &
\n[0]{tqdm/DStar/@X}    &
\n[0]{tqdm/Ochiai/@X}     &
\n[0]{tqdm/Tarantula/@X}        &
\n[0]{tqdm/ST/@X} &
\n[0]{tqdm/MBFL/time}   &
\n[0]{tqdm/PS/time} &
\n[0]{tqdm/SBFL/time}  &
\n[0]{tqdm/ST/time} \\

\project{youtube-dl}     &
$124.5$  & $\numprint{2367}$   & $16$   &
\n[0]{youtube_dl/Metallaxis/@X}         &
\n[0]{youtube_dl/Muse/@X}   &
\n[0]{youtube_dl/PS/@X} &
\n[0]{youtube_dl/DStar/@X}    &
\n[0]{youtube_dl/Ochiai/@X}     &
\n[0]{youtube_dl/Tarantula/@X}        &
\n[0]{youtube_dl/ST/@X} &
\n[0]{youtube_dl/MBFL/time}   &
\n[0]{youtube_dl/PS/time} &
\n[0]{youtube_dl/SBFL/time}  &
\n[0]{youtube_dl/ST/time} \\

\cmidrule{2-15}
\multicolumn{1}{r}{\textbf{total}} &
$1253.5$ &
$\numprint{112602}$ &
$135$ &
\n[0]{all/Metallaxis/@X}         &
\n[0]{all/Muse/@X}   &
\n[0]{all/PS/@X} &
\n[0]{all/DStar/@X}    &
\n[0]{all/Ochiai/@X}     &
\n[0]{all/Tarantula/@X}        &
\n[0]{all/ST/@X} &
\n[0]{all/MBFL/time}   &
\n[0]{all/PS/time} &
\n[0]{all/SBFL/time}  &
\n[0]{all/ST/time} \\

\bottomrule
\end{tabular}
\label{tab:bugsinpy-projects}
\end{table*}

\section{Experiments}
\label{sec:experiments}

To demonstrate \fp's applicability
to projects of realistic size and complexity,
\autoref{tab:bugsinpy-projects} summarizes
a few key results
of the large-scale experiments we performed in our related work~\cite{Rezaalipour:2023}.
These experiments involved
135 bugs from 13 open-source Python projects
taken from the \bip curated collection
of real-world Python bugs~\cite{Widyasari:2020}.
Each bug $b$ in \bip consists of
two revisions $B_b, F_b$ of a Python project complete with
its programmer-written tests;
the first revision $B_b$ includes a bug exposed by the tests,
and the second revision $F_b$ is the programmer-written fix.
Overall, these experiments involve over 1.2 million lines of code
and over 100 thousand tests.

For each bug $b$,
we ran \fp on each buggy revision $B_b$,
and used the fixed revision $F_b$ to determine
whether \fp localized
the actual bug locations (i.e., where the programmer edited the program to fix it).
As key metric of fault localization accuracy (effectiveness),
\autoref{tab:bugsinpy-projects} reports the $@5$ count
for each fault localization technique:
the number of bugs that the technique correctly localized
within the top-5 ranks of its output.
This is a common metric of fault localization effectiveness,
which is based on a scenario
where the user only inspects a few (i.e., five)
locations in the output, and ignores any other locations that are ranked lower.
\autoref{tab:bugsinpy-projects}
indicates that SBFL techniques (DStar, Ochiai, Tarantula)
are the most effective ones according to this metric,
followed by MBFL techniques (Metallaxis, Muse).
As we explained intuitively in \autoref{sec:using},
PS and ST are more specialized techniques
that are only applicable to bugs that involve branching predicates (PS)
or that result in a crash (ST);
in fact, they are accurate only for a fraction of the bugs in the experiments.

The average running time of \fp
on each bug (also reported in \autoref{tab:bugsinpy-projects})
confirms on a much larger scale
the same trends that \autoref{sec:using}'s
toy examples demonstrated in the small.
Namely, ST is by far the fastest technique,
since it just runs failing tests
(usually, only a handful of a whole test suite);
SBFL is still nimble but has to run \emph{all}
tests while collecting coverage information;
PS and MBFL take considerably more time,
since they have to run all tests
on several variants of the programs.
We refer interested readers to our related work~\cite{Rezaalipour:2023}
for many more details about the practical capabilities
of different fault localization techniques on Python programs.

\section{Related Work}
\label{sec:related-work}

Fault localization research spans over three decades~\cite{Wong:2016},
during which it produced
diverse fault localization techniques,
using a variety of sources of information
(e.g., traces, mutation analysis, static analysis)
as heuristics to
identify locations suspicious of being implicated with a fault.
For lack of space, we only outline some essential related work,
and refer readers to surveys for more details and references~\cite{Wong:2016}.

Spectrum-based fault localization (SBFL)~\cite{Wong:2014,Abreu:2007,Jones:2005}%
---based on a straightforward dynamic analysis of traces---%
is probably the most studied family of techniques.
Despite the apparent simplicity of SBFL techniques,
they remain generally quite effective and
scalable~\cite{Zou:2021,Rezaalipour:2023}.
In contrast, mutation-based fault localization (MBFL)~\cite{Moon:2014,Papadakis:2015}
computes a location's suspiciousness using mutation analysis
(i.e., the number of tests that ``kill'' a mutant that targets the location).
MBFL has gained attraction in recent years,
since it can outperform SBFL for certain categories of faults;
however, it remains a computational expensive approach,
since the more mutants are generated the more test executions are needed.

Practical adoption of fault localization techniques require scalable fault localization tools that work on real-world programs. Although
there exists several fault localization tools~\cite{Janssen:2009,Ribeiro:2018,Horvath:2020,Chen:2016} in the literature, they are mostly developed
for programming languages such as Java, C/C++, and the .NET languages. 
Despite Python's popularity, there is not much fault localization research targeting Python~\cite{Widyasari:2022,Rezaalipour:2023};
and even less tool support~\cite{Sarhan:2022}.
To our knowledge,
CharmFL~\cite{Sarhan:2021}
is the only publicly available fault localization tool for Python
other than \fp.
CharmFL, implemented as a plugin of the PyCharm IDE,
only supports SBFL techniques.
In contrast, \fp supports a variety of techniques and granularities,
which was instrumental to perform our large-scale empirical study
of fault localization in Python~\cite{Rezaalipour:2023}.

\section{Conclusions and Tool Availability}
\label{sec:availability}
\label{sec:conclusions}

This paper presented \fp, an automated fault localization tool for Python programs. We explained the motivation behind developing \fp, its implementation details, and simple examples of usage. We also summarized the experimental results of running \fp on 135 bugs in 13 popular real-world projects,
which demonstrate that it is a flexible tool, usable on projects of considerable size.

Users can easily install \fp from PyPI\footnote{\url{https://pypi.org/project/fauxpy}}, using \Bash{pip install fauxpy}.
\fp's source code
is also publicly available.\footnote{\url{https://github.com/atom-sw/fauxpy}}
A
companion repository\footnote{\url{https://github.com/atom-sw/fauxpy-experiments}}
makes available the complete dataset of our related empirical study~\cite{Rezaalipour:2023}.

\bibliographystyle{IEEEtran}
\bibliography{IEEEabrv,main-bib}

\end{document}